# Acoustic Wave Approach for Multi-Touch Tactile Sensing

*Yuan LIU[*1,2], Jean-Pierre Nikolovski[1], Moustapha Hafez[1], Nazih Mechbal[2], and Michel Vergé[2]*
[1] CEA, LIST, Sensory Interfaces Laboratory, 18 Route du Panorama, 92265 Fontenay-aux-Roses, France
[2] LMSP (CNRS 8106), Arts et Métiers ParisTech, 151 Boulevard de l'Hôpital - 75013 Paris, France

**Abstract:**
**In this communication, we present a high resolution tactile plate that can localize one or two contact fingers. The localization principle is based on Lamb wave absorption. Fingers' contact will generate absorption signals while Lamb waves are propagating in a thin finite copper plate. These signals can be related to the contact positions and can be calibrated before the use of tactile plate. Fingers' contact positions are calculated by finding the closest calibration signal to the measured signal. Positions are carried out in less than 10 ms with a spatial resolution of 2 mm for one finger localization. Multi-points localization by this technology is developed and a two-point case is initialized and tested. Several optimization methods are also presented in this paper, as the double validation check which could improve the accuracy of single-point localization from 94.63% to 99.5%.[*]**

## 1. INTRODUCTION

Compare to other technologies of tactile sensing [1]-[4]. Acoustic wave methods have many advantages as these processes don't require an extra layer, and they could be implanted on surface with different materials, such as glasses or copper.

Lamb wave methods have been used for the contact sensing between an index (pinpoint or finger) and a plate [5]-[7]. Localization may depends on a time delay of an acoustic wave propagating in the object or on a cross correlation with a predefined acoustic signature associated with the position.

The time delay method requires a complex real-time circuit board. It would be difficult to distinguish the delay time of original waves generated by fingers' contact at different positions from those of reflection waves generated by the plate boundaries.

For the cross correlation method, acoustic signature is established on the absorption of Lamb wave while an index is in contact with the plate. The difficulty for multi-touch is that there isn't the correspondence between acoustic signature associated with multi-contact position combination and those signatures associated with single contact position.

Thus, this paper proposes a new method using Lamb wave absorption for multi-points tactile application.

A copper plate with propagating Lamb wave contains plural predefined discrete points (X × Y) on the surface. One calibrated reference vector of Lamb wave is associated with one discrete point (x, y), and it could be considered as one pattern. When a subject is in contact with this plate, we could measure a modified vector of the propagating Lamb wave, which could be considered as an object. The localization could be then described as a pattern recognition process.

The multi-point localization is made by the same way, while the predefined pattern is not associated with one point on the tactile surface, but a combination of several predefined points.

## 2. EXPERIMENTAL

### 2.1 Test Bench

The Lamb wave device is integrated in a test bench which is described in Fig.1. It contains a function generator Tektronix 3012 and a data acquisition board Pico ADC 212/50. Both instruments are connected to a laptop PC by USB ports. Matlab is used for the signal processing and demonstration of results.

Four piezo-ceramic transducers (PZT, thickness = 0.5 mm, type Pz27 by Ferroperm, Denmark) are bonded to the edges of the plate with a conductive glue Circuitworks CW2400. The transducers are used as transmitters and receivers of Lamb wave. The transducers are placed to break the symmetry of plate, as shown in Fig.2.

The Data acquisition board is set at 195.31 kHz as sampling frequency. A 2 ms acquisition collects 390 points values as signals' amplitudes at different frequency component. The FFT results contain information of signals amplitudes from 0 to 97.7 kHz. In the localization process, we take into account not only the amplitude of signals at 30 selected frequencies of excitation signal, but the full frequency.

---

[*] Corresponding author: Email: yuan.liu@cea.fr or 2006-d113@etudiants.ensam.fr

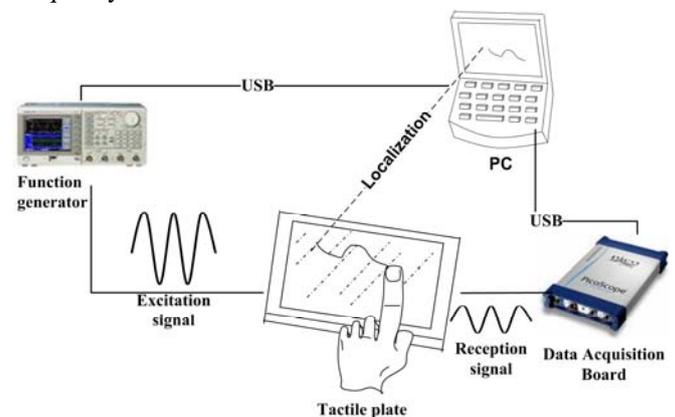

Fig. 1. Test bench with a Lamb wave plate.

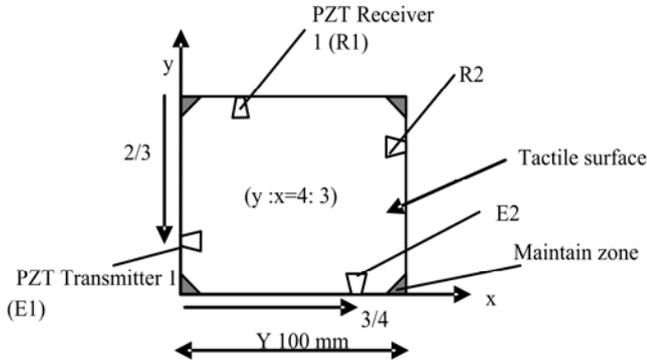

Fig. 2. Placement of the PZT transducers on the rectangular plate (75 mm × 100 mm).

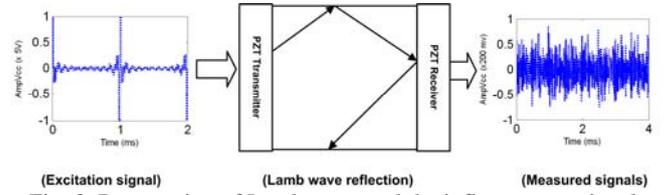

Fig. 3. Propagation of Lamb wave and the influences on signals

By analytical results [8]-[9], a Lamb wave can reflect 23 times at the boundary in 1 ms, when its propagating velocity is equal to 2270 m/s and the dimension of the plate is 75 mm × 100 mm.

Based on a simulation study of Lamb wave propagation [8], in which we found that the board conditions will modify the Lamb wave propagation, a support is designed to minimize and control the influence of boundary condition. The 75 mm × 100 mm rectangular thin finite copper plate (thickness = 450 μm) is maintained at the four vertexes by the support.

The test bench works as described in Fig.3. The excitation signal is sent from the function generator to the PZT transducers, which will generate Lamb waves. After short-time stabilization, the interaction between the original wave generated by the PZT transmitter and the reflection waves produced by the board of the interface establishes a static Lamb wave figure. Then the PZT Receiver acquires the static signals and transfers them to computer for proceeding.

A sinusoidal excitation signal could only provide single amplitude attenuation information produced by contact at different position. It will not be enough for distinguishing the numerous contact positions, who would have same attenuation amplitude at one frequency.

Lamb waves at different frequencies have different propagating velocities and ratios between the normal and the shear components, but they have different attenuation behaviors when a finger in contact with the plate. The tactile localization by Lamb wave is an extraction of position information from different attenuated Lamb wave signals.

To have a more precise localization, more different and well spaced frequencies are required. The more frequencies we use, the more information we could gather from their absorption comportments. However, those frequencies should be well spaced enough to avoid any confusion in fast discrete Fourier transform (FFT) for signal proceeding. There is a compromise to make. The excitation signal is then composed of 30 frequencies distributed from 20 kHz to 80 kHz, with an interval of 2 kHz.

## 2.2 Localization Principle

Before the use of tactile plate, we need a calibration process to collect references contact positions. Pixilation with discrete points is used for addressing the tactile surface. As a first test, we choose a mesh with 5 lines and 5 columns. Each mesh element is 15 mm × 20 mm. The predefined positions of contact are the centre of mesh, and the centre of mesh boundaries. The number of calibration reaches 81, with a spatial resolution of 7.5 mm × 10 mm.

At each reference position, a subject put one finger with a contact surface approximately 1 cm$^2$. Then we measure the attenuated signals from two PZT Receivers independently. The FFT figure of this attenuated signal is then considered as a pattern. It could be descried as:

$$P_{mref} = (A_{1mref}, A_{2mref}, \ldots A_{imref}, \ldots A_{Nmref}) \quad (1)$$

Here $m$ is the index of Lamb wave receiver, $A_{imref}$ is the amplitude of one frequency element of the FFT figure, $N$ is one half of the acquisition points acquired by the Data acquisition board.

All 81 predefined positions are numerated and arranged in an array indexed from $(x_1, y_1)$ to $(x_9, y_9)$.

Once 81 reference FFT figures are collected by the calibration process, they are considered as 81 patterns associated with predefined contact positions. In the localization process, a subject (may not be the same as in the calibration step) put his finger at one reference point, a FFT figure of attenuated signal is stored as an object, which could be described as a vector:

$$P_m = (A_{1m}, A_{2m}, ... A_{im}, ... A_{Nm}). \qquad (2)$$

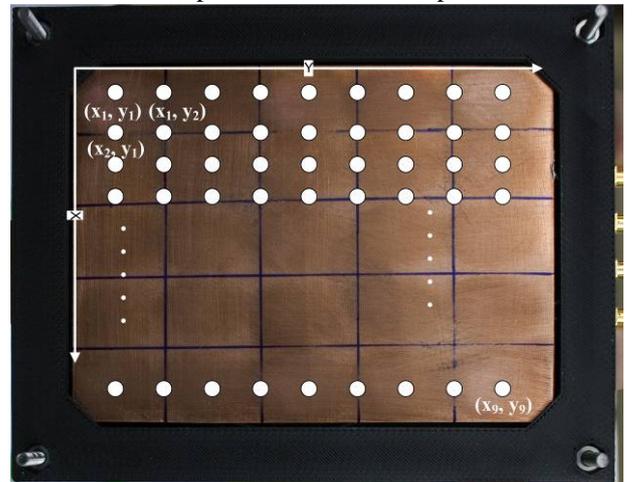

Fig.3. Pixilation of plate.

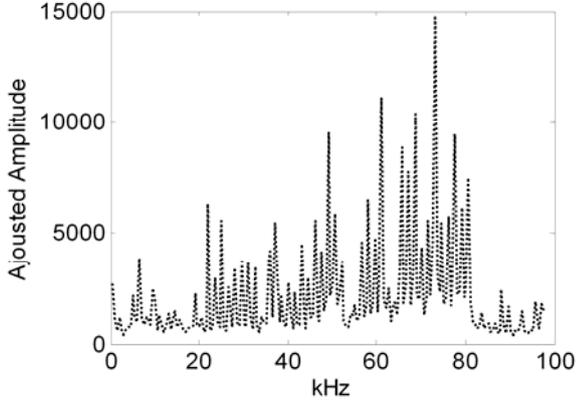

Fig.4.  FFT figure of acquired signals. A human finger is in contact at the position $(x_1, y_1)$, by PZT Receiver 1.

Localization of the contact position is to find out the closest pattern to the object. The nearest neighborhood search is used with different definitions of distance (dm), such as Manhattan Distance:

$$dm(Pm, Pmref) = \sum_{i=1}^{n} |A_i - Aref_i| \qquad (3)$$

The calibration by human finger has much incertitude in localization, since the finger of each person is unique with its specials surface properties (fingertips), and different Young's module. Figure 5 show 10 000 consecutive localization results when a human finger is put at the position $(x_5, y_5)$. Lamb wave receiver R1 has 7592 correct measurements, but R2 has only 1438 measurements correspond to the correct position, even less than a wrong position $(x_4, y_3)$ (2435 of 10000 measurements). It is because some contact positions are difficult to provide attenuated information for some Lamb wave receiver. These positions could be considered as blind area for one receiver. The $(x_5, y_5)$ is difficult for Lamb wave receiver R2 to detect, but the same receiver has a good precision for localization the $(x_3, y_7)$ point while R1 has a less good accuracy.

## 2.3 Multi-Touch Extension

The tactile plate developed in this paper could be used for multi-points tactile localization with simply extension of the same analysis on attenuated signals as that for one point contact sensing. For the localization of one contact point, each reference signal is associated with a predefined contact point; in multi-points case, reference signal is associated with a combination of plural points. For example in 2 simultaneous contact points localization, a reference attenuated signal could signify a 4-D point (x,y,x',y') where (x,y) the position of the first contact point and (x',y') is the position of the second.

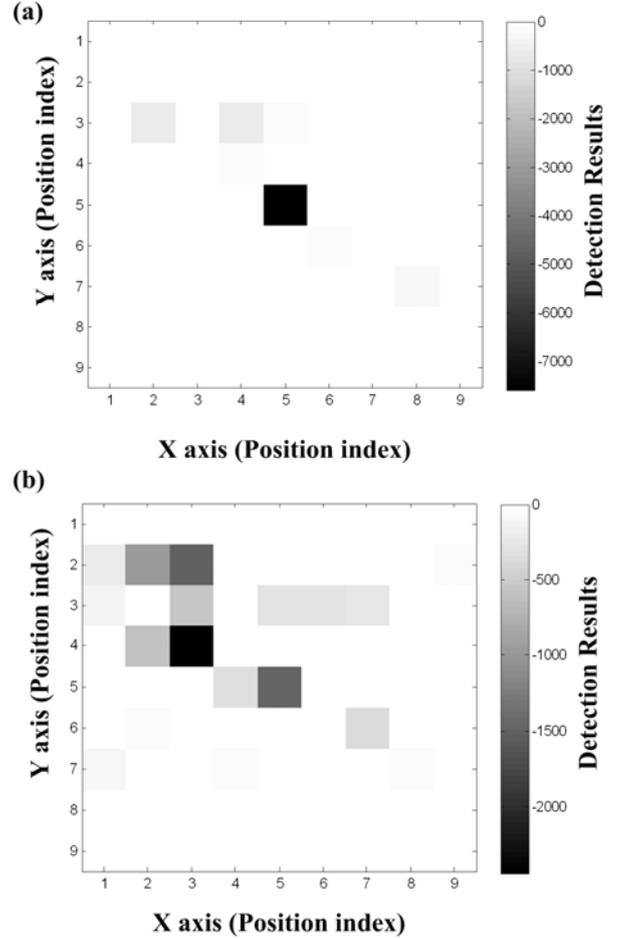

Fig. 5. (a)  10000 consecutive measure results by Lamb wave receiver R1 while a finger is putting at position $(x_5, y_5)$. (b) Measure results by R2.

In the single-point localization, the measurement results could be illustrated directly by a contrast figure of Manhattan distance. It is difficult to illustrate a multiple dimension points for multi-points localization. In this paper, we arranged all the reference signals in an array.  For example, as the plate is pixilated with 5×5 contact points, $P_1$ means first configuration as $(x_1,y_1,x'_2,y'_2)$ where 2 fingers are on the   $(x_1,y_1)$ and $(x_2,y_2)$ positions respectively.

Considering the two fingers used in the calibration process are similar, $(x_1,y_1,x'_2,y'_2)$   and $(x_2,y_2,x'_1,y'_1)$ means the same position combination of two-contact points. We impose in our system that the index of the second position is always higher than the first one, to simplify the calculation. All of possible two-point combinations of contact position are calibrated to collect a 300 length array defined by $P_{1-100}$ in order, as $P_2$ $(x_1,y_1,x'_2,y'_3)$, $P_{24}$ $(x_1,y_1,x'_5,y'_5)$ and $P_{300}$ $(x_5,y_4,x'_5,y'_5)$.

In some current methods of multi-touch applications, thumbs are considered as different from fingers, it could be taken into account in calibration process by modifying the indexation of reference signals.

The number of reference signals depends on the pixilation degree of the plate. For a plate with a 5×5 mesh, the number of potential combinations is $C_{25}^2$ or 300. In extension, with M predefined points on the surface and N multiple contact fingers, this number is $C_M^N$.

Fig.6 shows a valid measurement with both Lamb wave receivers. The acquired signal has minimum distance to the reference signals numerated 194 by both receivers, which is the index for $(x_2,y_5,x'_5,y'_4)$. Two fingers are detected at position (x=30mm, y=100mm) and (x=75mm, y=80mm), respectively.

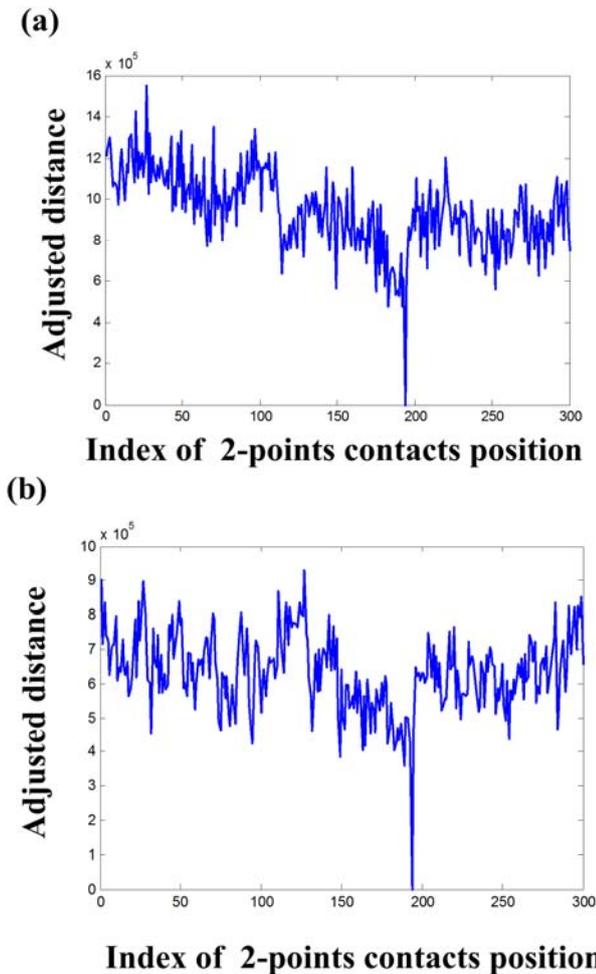

Fig.6. (a) Adjusted distance from one acquired signal to references signals, by Lamb wave receiver R1. (b) Adjusted distance from one acquired signal to references signals, by Lamb wave receiver R2.

## 3. OPTIMISATIONS

To improve the accuracy of localization, we propose two methods of optimisation. The first is a double validation check to combine and coordinate the sensing capacity of two Lamb wave receivers; the second is the use of artificial finger in the calibration process.

### 3.1 Double Validation Check

The double-validation check used in this paper means a discrimination of measures results. One measure of contact position is valid if and only if the closest reference signals found by both two receivers independently indicate the same position. Fig.7. shows the localization results after the double-validation check, when a human finger is put at the position $(x_5, y_5)$. It is the optimization results of Fig.5. There is only one neighbouring error $(x_5, y_4)$ among 1369 valid measurement, the localization is correct with 99.93%.

However, this check will slow up the system. As the acquisition time is limited by 2 ms. The response time of measurement in the whole loop is approximately 3 ms with the calculate of 81 distances and finding the minimum one. The fact that only 14 % measurement could be valid means the average response time is prolonging to about 20 ms.

Some others positions such as $(x_3, y_7)$ have better results. We could have 7424 valid measurement in 10000 consecutive measures with 99.49% accuracy, as shown in Fig.8. The average response time in this case is about 5 ms.

If we consider the neighbouring points are correct localizations, the accuracy of measure is 100% in both two cases.

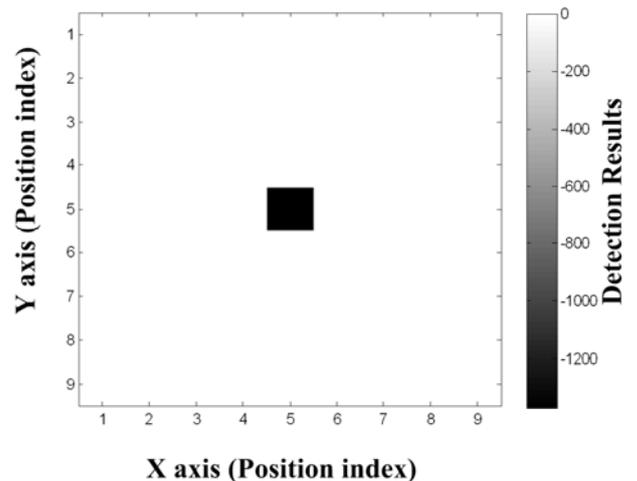

Fig. 7. Valid localization results after double-validation check, while a finger is putting at position $(x_5, y_5)$.

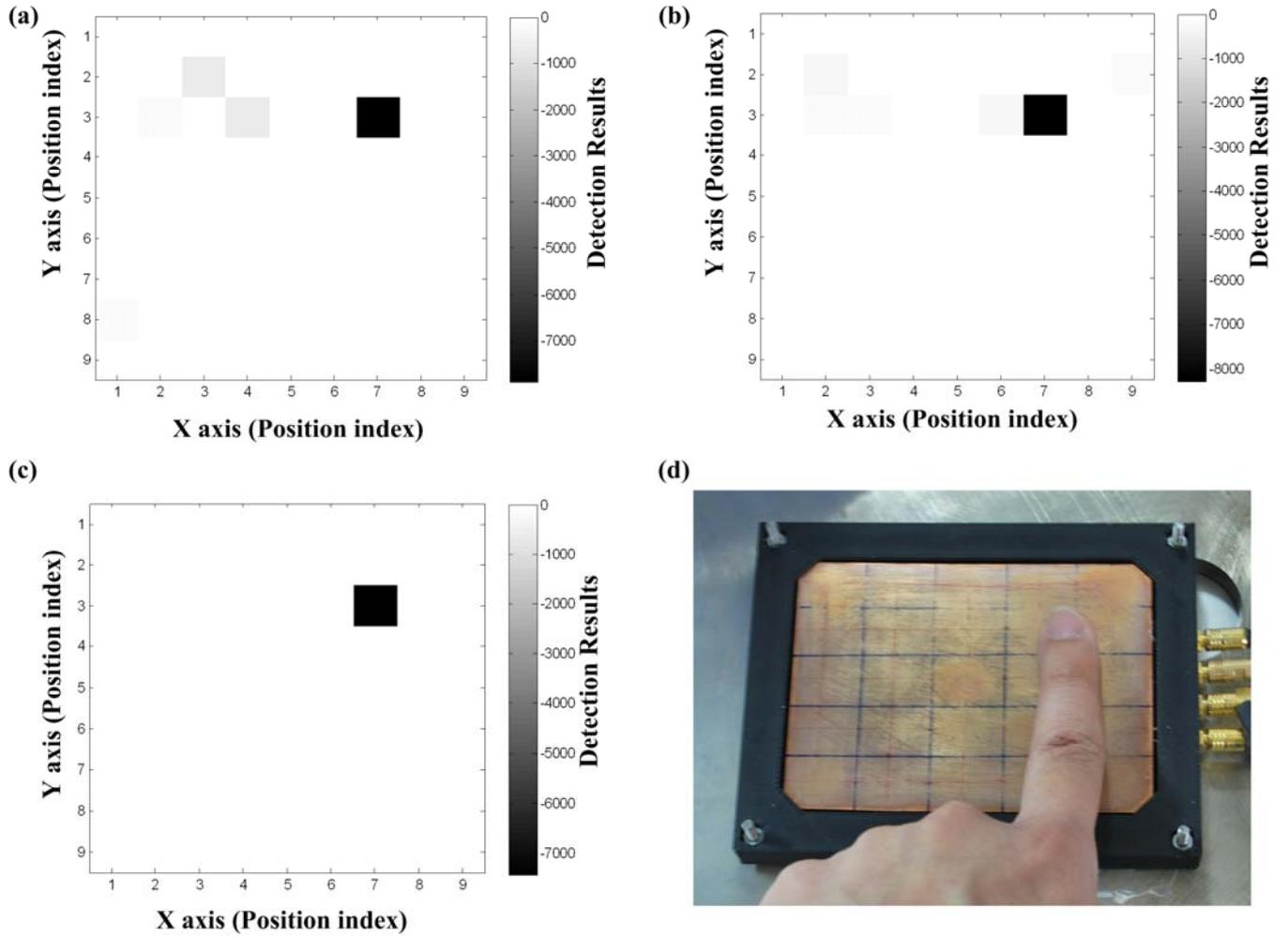

Fig. 8. (a) 10000 consecutive measure results by Lamb wave receiver R1 while a finger is putting at position $(x_3, y_7)$. (b) Measure results by R2. (c) Valid measurements after double-validation check. (d) Photo of finger contact at $(x_3, y_7)$.

## 3.2 Silicone Finger Calibration

Human finger calibration has a good result in localization after double validation check, but it is difficult to improve the spatial resolution of the tactile surface, since it is difficult for a volunteer to hold a small interval between two predefined contact points such as 3 mm with a good precision. For the plate dimension used in this paper, a calibration process for a 3 mm ×3 mm resolution requires up to 825 references points. An artificial finger made with silicone is necessary.

The design of artificial finger is not discussed in this paper. A detailed study of silicone prosperities and difference from human finger has been presented in [10].

As we have already a matrix of reference attenuated signals by human finger contacts, the artificial finger is arranged and modified until the tactile plate could find its position on the surface.

The silicone finger is taken by a 3-D robot with a 2 μm precision. The resolution is set to 32 × 36 contact points on the tactile surface, with a resolution spatial 2 mm × 2.5 mm. The tactile surface is 64 mm × 90 mm. Fig.9. present the artificial finger in contact with the tactile plate.

After the calibration process, a subject put one finger at the center of tactile plate, and we observe the 10000 consecutive measure results. By Lamb wave receiver (R1), 94.63% measurements localize the contact point to a same reference points $(x_{18}, y_{20})$ which means the position (x=36 mm, y= 50 mm). The errors could come from the pulse of human. By this consideration, double-validation check is not necessary to improve the accuracy of localization; other

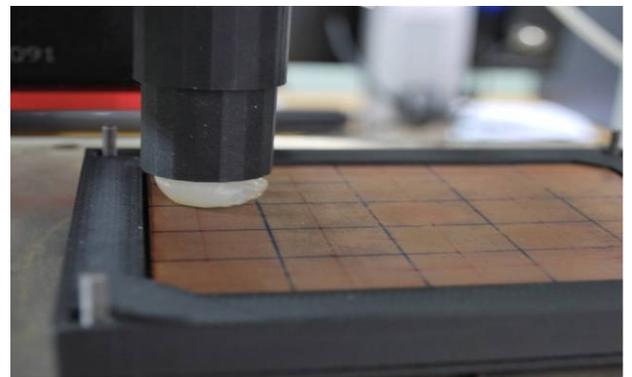

Fig.9. Silicone finger's contact on the tactile surface. It is characterized till it could localized by reference signals by human finger calibration.

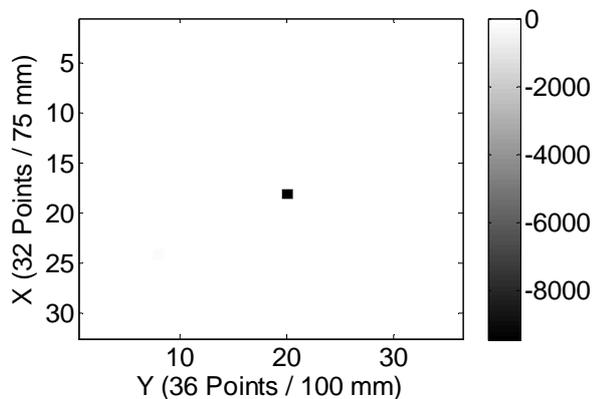

Fig.10. Localization result by R1, contrast image of 10000 consecutive measurements.

statistic method is under studying.

The use of artificial finger improves the robustness of the system, as the calibration is more precise in position, in contact surface and in contact pressure, etc. As the double-validation check will not be needed, it could also improve the response time.

As the tactile surface is discrete, when a human finger contact at the middle of two predefined contact points, it will have some ambiguity of measurement. As shown in Fig.11, when a contact point at the top right of the plate, two neighboring points are indicated by the localization process, 890 measures detect $(x_8, y_{27})$ and 110 detect $(x_8, y_{28})$. However, it could be both considered as correct measures, with 100 % accuracy.

## 4. CONCLUSIONS AND PERSPECTIVES

This paper proposes and develops a tactile plate with simple architecture. The tactile plate has a spatial resolution up to 2 mm with a response time at millisecond level. With a double-validation process, simple point localization has a good accuracy. Multi-points localization by this technology is also developed and a two-point case is initialized and tested.

In the next future, the number of reference points will be increased to develop a quasi-continue tactile plate, which requires solving the problem of ambiguity between proximate points.

Though the algorithm used in this work is just a simple approach to pattern recognition, modern algorithms such as on-line neural network optimization is also being studied, expecting the reduce of computation and response time.

Circuit board for mobile application, which should replace the function generator and data acquisition board, is also in progression.

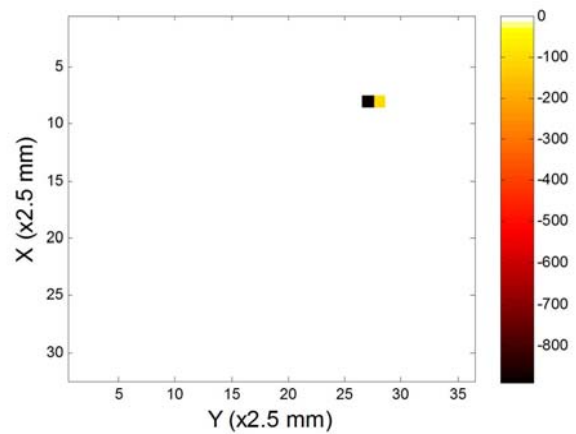

Fig.11. 1000 consecutive localization results when a finger contact is produced. This figure shows the ambiguity between two predefined points.

# Acoustic Wave Approach for Multi-Touch Tactile Sensing

Y. LIU*[1,2],   JP. Nikolovski[1], M. Hafez[1] , N. Mechbal[2], and M. Vergé[2]

[1] CEA, LIST, Sensory Interfaces Laboratory, 18 Route du Panorama, 92265 Fontenay-aux-Roses, France

[2] LMSP (CNRS 8106), Arts et Métiers   ParisTech,   151 Boulevard de l'Hôpital - 75013 Paris, France

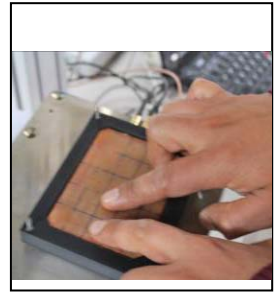

In this communication, we present a high resolution tactile plate that can localize one or two contact fingers. The localization principle is based on Lamb wave absorption. Fingers' contact will generate absorption signals while Lamb waves are propagating in a thin finite copper plate. These signals can be related to the contact positions and can be calibrated before the use of tactile plate. Fingers' contact positions are calculated by finding the closest calibration signal to the measured signal.  Positions are carried out in less than 10 ms with a spatial resolution of 2 mm for one finger localization. Multi-points localization by this technology is developed and a two-point case is initialized and tested. Several optimization methods are also presented in this paper, as the double validation check which could improve the accuracy of single-point localization from 94.63% to 99.5%. In the next future, the number of reference points will be increased to develop a quasi-continue tactile plate, which requires solving the problem of ambiguity between proximate points. Though the algorithm used in this work is just a simple approach to pattern recognition, modern algorithms such as on-line neural network optimization is also being studied, expecting the reduce of computation and response time. Circuit board for mobile application, which should replace the function generator and data acquisition board, is also in progression.